\documentclass[10pt,numbers]{sigplanconf}

\usepackage[T1]{fontenc}
\usepackage{microtype}
\usepackage{flushend}

\usepackage{amsmath}

\usepackage[pdftex]{graphicx}
\graphicspath{{./figures/}}
\DeclareGraphicsExtensions{.pdf}

\usepackage[hyphens]{url}

\begin{document}

\toappear{}

\setlength{\pdfpageheight}{\paperheight}
\setlength{\pdfpagewidth}{\paperwidth}

\conferenceinfo{CONF 'yy}{Month d--d, 20yy, City, ST, Country}
\copyrightyear{20yy}
\copyrightdata{978-1-nnnn-nnnn-n/yy/mm}
\copyrightdoi{nnnnnnn.nnnnnnn}


\titlebanner{}        
\preprintfooter{}     

\title{A Quantitative Study of Java Software Buildability}
\subtitle{}

\authorinfo{Mat\'u\v{s} Sul\'ir\: and \:Jaroslav Porub\"an}
           {Department of Computers and Informatics\\
            Faculty of Electrical Engineering and Informatics\\
            Technical University of Ko\v{s}ice\\
            Letn\'a 9, 042 00 Ko\v{s}ice, Slovakia}
           {\{matus.sulir,jaroslav.poruban\}@tuke.sk}

\toappear{
\fontsize{7pt}{7pt}\selectfont
\copyright{} Mat\'u\v{s} Sul\'ir and Jaroslav Porub\"an 2016. Publication rights licensed to ACM.
\\
This is the author's version of the work. It is posted here for your personal use. Not for redistribution. The definitive version was published in the following publication:\\
\\
\fontsize{8pt}{8pt}\selectfont
\textit{PLATEAU'16}, November 1, 2016, Amsterdam, Netherlands\\
ACM. 978-1-4503-4638-2/16/11...\\
http://dx.doi.org/10.1145/3001878.3001882
}

\maketitle

\begin{abstract}
Researchers, students and practitioners often encounter a situation when the build process of a third-party software system fails. In this paper, we aim to confirm this observation present mainly as anecdotal evidence so far. Using a virtual environment simulating a programmer's one, we try to fully automatically build target archives from the source code of over 7,200 open source Java projects. We found that more than 38\% of builds ended in failure. Build log analysis reveals the largest portion of errors are dependency-related. We also conduct an association study of factors affecting build success.
\end{abstract}




\section{Introduction}

In the broadest sense, the purpose of a build system is to generate output data given a set of inputs \cite{Smith11software}. Considering a more specific, typical scenario, a build system reads a buildfile -- e.g., a Maven POM (Project Object Model) file. Using the supplied information, it downloads necessary third-party components. Next, it executes a compiler to produce binary files from textual source code. Finally, it packages the program in a format suitable for deployment.

As tools like Make have existed since the 70's \cite{Feldman79make}, build systems are often perceived as a solved problem by researchers \cite{Neitsch12build}. Nevertheless, developers often encounter build problems which negatively affect their work \cite{Kerzazi14why}.

A software system should be buildable from source in one step \cite{Spolsky04joel}. In the Java ecosystem, we expect an invocation of a build system like Maven, Gradle, or Ant to represent this step.

In this paper, we aim to provide empirical evidence that build system invocations of many open source projects fail. Furthermore, we will look at the reasons of failures and factors affecting them.

\subsection{Motivation}

Consider the following three situations:

\begin{itemize}
\item A researcher wants to use an open source system in his next program comprehension experiment.\footnote{We are motivated by our frustration when we tried to build multiple large open source projects without success. This also partially affected some decisions in our experiments and studies \cite{Sulir15sharing,Sulir16locating}.} He downloads the software project of interest in a form of a source code archive, opens it in an IDE (integrated development environment) and tries to build it.

\item A student is willing to contribute to her favorite open source program with a new feature. She pulls the source from the project's version control system. Before starting the actual work, she checks whether the project can be built using a command-line interface of the build system.

\item A practitioner is trying to fix a bug in a third-party library their company extensively uses. Again, the necessary precondition is to make sure the library builds.
\end{itemize}

All three situations have something in common: The message ``Build failed'' is likely to appear. Instead of doing useful work, the developers start to inspect cryptic error messages, search them on the internet, modify project configuration files, manually install packages and configure paths. This can take hours to fix -- or, at worst, the developer gives up. Such situations significantly hinder the usability of build systems.

\subsection{Aim}

While there exists anecdotal evidence that open source systems are sometimes difficult to build \cite{Neitsch12build}, large-scale studies on a substantial number of projects are rare.

We strive to simulate a simple programming environment of a Java developer, download a large number of open source Java projects from the software forge GitHub, fully automatically execute the build command for each of them, and observe the build process outcomes.

We formulate our research questions as follows:
\begin{itemize}
\item \textbf{RQ1:} What portion of projects fail to build?
\item \textbf{RQ2:} What types of build errors do occur most frequently?
\item \textbf{RQ3:} Is there association between the build success/failure and other project's properties (e.g., the used build tool, age)?
\end{itemize}

\subsection{Methodology}

\begin{figure}
\centering
\includegraphics[width=0.9\linewidth]{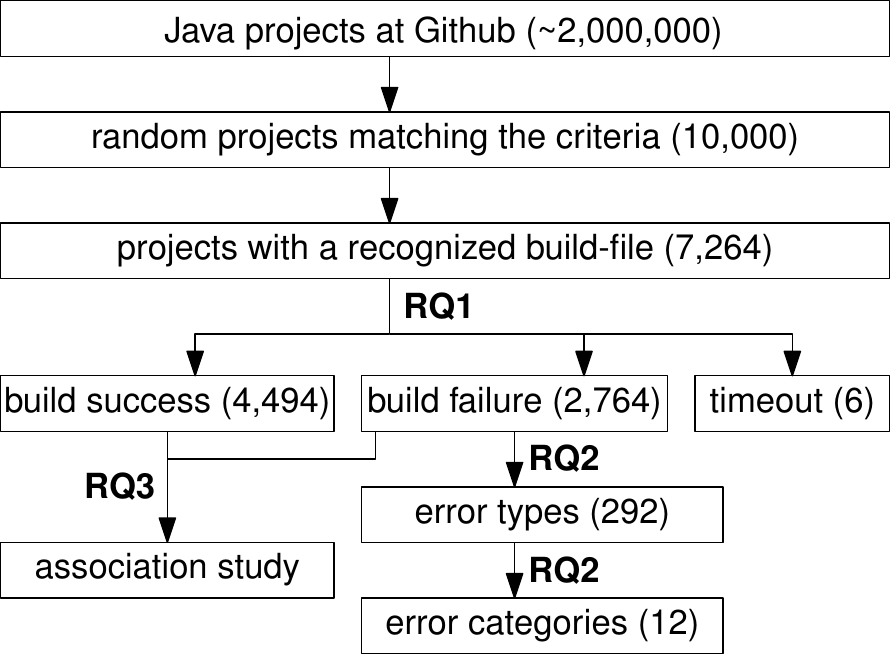}
\caption{The study overview}
\label{f:overview}
\end{figure}

For a high-level overview, see Figure~\ref{f:overview}. In section~\ref{s:proportion}, we answer \textbf{RQ1} by a simulation study. Using the collected data, a semi-automated classification of failed builds (\textbf{RQ2}) is presented in section~\ref{s:types}. An association study is also performed on the data (\textbf{RQ3}), which is described in section~\ref{s:relation}.

\section{Build Failure Proportion}
\label{s:proportion}

In this section, we answer the first research question.

\subsection{Method}

The data collection process consisted of two steps: obtaining a list of suitable projects, and the building process itself. The whole process was fully automated, controlled by a script.

\subsubsection{Project Selection}

From a set of all public repositories at a large software forge, GitHub, we selected a random sample of projects corresponding to the following inclusion criteria:
\begin{itemize}
\item projects written in Java,
\item not using JNI (Java Native Interface), Android or Java Micro Edition (ME) APIs,
\item having an open source license,
\item and forked at least once.
\end{itemize}

Now we will describe the rationale behind the criteria and details of the process.

In this study, we focused only on projects written in the Java language. Java seems like an ideal choice, since it is a compiled language and it offers very feature-rich build systems. To determine the project's main language, the GitHub API\footnote{\url{http://developer.github.com/v3/}} was used. There were approximately 2,000,000 Java projects on GitHub at the time the study was conducted (see Figure~\ref{f:overview}).

We were interested only in open source projects. To assess whether a particular project is open-source, we utilized the information supplied by the GitHub API. GitHub produces these data by matching the contents of ``License'' files with a set of well-known licenses. Around 20\% of GitHub repositories contain a recognized license.\footnote{\url{http://github.com/blog/1964-open-source-license-usage-on-github-com}}

Only repositories with at least one fork were included. First, this indicates that there is some interest in cooperation on a project. Second, this criterion lowered the number of results by a factor of 10 to keep the number of searched repositories reasonable.\footnote{The GitHub API does not support random selection of repositories, so we needed to retrieve metadata for all projects matching the criteria and sample them locally.}

We requested the GitHub API with a combination of queries corresponding to the above criteria. From the list of returned repositories, a random sample was selected. The most recent version (the current ``master'' branch) of each project was downloaded in a form of a tarball. The total number of downloaded archives was 14,567.

After extraction of the archives, the presence of JNI, Android and Java ME APIs was tested using file name and content patterns. The reason for exclusion of projects using these APIs was to maximize internal validity. Thanks to JNI, Java source code could call methods written in other languages, while we were interested in pure Java. We would be no longer testing just Java build systems, but also other toolchains outside the Java ecosystem, which also applies to Android and Java ME.

The resulting set of random projects, corresponding to the inclusion criteria, consists of 10,000 projects.

\subsubsection{Build Process}

\begin{table*}
\centering
\caption{The studied build systems (auxiliary log-formatting arguments were omitted)}
\label{t:build-systems}
\begin{tabular}{lll}
\hline
\textbf{Tool} & \textbf{File name} & \textbf{Build command} \\ \hline
Gradle & build.gradle & {\small\verb/gradle clean assemble/} \\
Maven & pom.xml & {\small\verb/mvn clean package -DskipTests/} \\
Ant & build.xml & {\small\verb/ant clean; ant jar || ant war || ant dist || ant/} \\ \hline
\end{tabular}
\end{table*}

For each of the 10,000 projects, a primary build tool was recognized by the script. In our study, we focused on three popular build systems: Gradle, Maven and Ant. To assess which build system a particular project uses, the presence of a build-file was checked in the project's root folder. For a list of file names, see Table~\ref{t:build-systems}, column ``File name''. In case multiple build-files are found in the root directory, Gradle takes precedence over Maven, and Maven has higher priority than Ant. The ordering is based on an assumption that newer systems are used as a primary build tool, while preserving the legacy ones for compatibility.

If a project contained build-file(s) only in non-root directories, or did not include a build-file at all, we excluded it from further processing. This leaves us with 7,264 projects.

The builds were executed using the lightweight virtualization platform Docker\footnote{\url{http://www.docker.com}}. The goal was to simulate a simple yet functional software environment of a Java programmer. The Docker image (available at \url{http://quay.io/sulir/builds}) contained the following software:
\begin{itemize}
\item the Linux distribution Fedora,
\item Java SE Development Kit (JDK) 8,
\item Gradle, Maven and Ant build tools,
\item the dependency manager Ivy,
\item the version control system Git,
\item a Ruby interpreter (to execute the control script),
\item and the ``tar'' and ``unzip'' utilities.
\end{itemize}

The control script executed the build process for each project sequentially. For every project, an appropriate build command was run, corresponding to the project's build tool -- see Table~\ref{t:build-systems}, column ``Build command''. Their general goal is to produce a runnable Java archive from source files. The underlying idea is that a developer should be able to execute the corresponding command on any project using the given build tool, without reading and following any complicated instructions or modifying the project's files, and the build output should be generated successfully. In cases when the build tool enables exclusion of tests via command-line arguments, we utilized it, since the output archive can be generated and software can be often successfully executed even if tests fail.

During each build process execution, both standard and error output streams of a build process were redirected to a log file. The exit code of the process was recorded: a non-zero exit code signifies failure (or timeout). The execution time of one build was limited to one hour. This was necessary since we encountered infinite builds during pilot runs. The project data -- exit codes and auxiliary metrics like the number of files in source archives -- were stored in a CSV (comma-separated values) file for further analysis.

\subsection{Results}

\begin{figure}
\centering
\includegraphics[width=\linewidth]{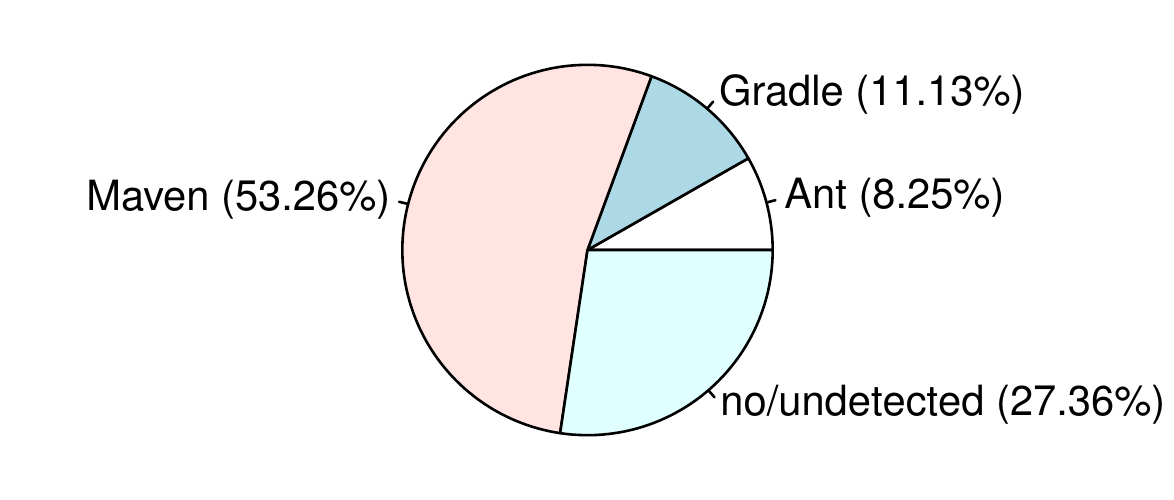}
\caption{The recognized build tools}
\label{f:tools}
\end{figure}

In total, 72.64\% of 10,000 projects have a recognized buildfile in its root directory. The most popular tool is Maven -- see Figure~\ref{f:tools}.

Among other findings, 8.54\% of projects contained such buildfile in a non-root directory. Since Ant does not provide dependency resolution capabilities itself, we were interested in how many projects using it also utilize the dependency manager Ivy. The number is relatively low: 10.18\%.

\begin{figure}
\centering
\includegraphics[width=\linewidth]{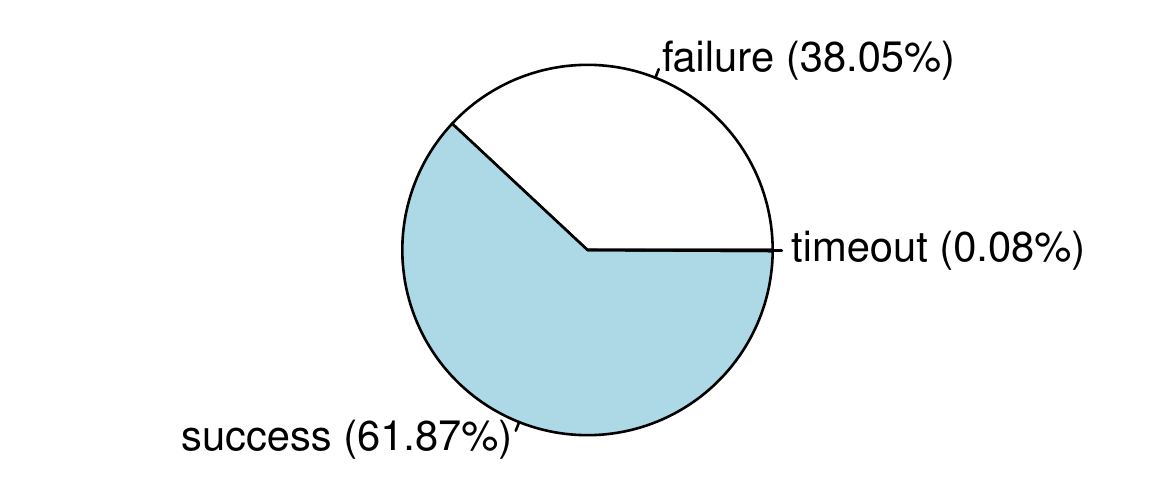}
\caption{The build success vs. failure}
\label{f:status}
\end{figure}

In Figure~\ref{f:status}, we see that 38.05\% of builds failed. This means in almost 4 of 10 cases, the programmer would not be able to produce a target archive from the source code without manual intervention.

A negligible part of builds did not meet the time requirement. Some of the reasons are lengthy dependency resolution and downloading, and expecting user input.

\section{Build Error Types}
\label{s:types}

In this section, we will analyze the reasons of the mentioned failures.

\subsection{Method}

To determine the kind of failure, the logs were searched for the presence of certain patterns. Since each build tool has its own log format, first we associated a buildtool-dependent, machine-readable error \textit{type} with each log (e.g., ``MojoFailure:maven-compiler-plugin''). This process was automated. Second, a significant portion of error types was manually grouped into tool-independent, human-readable \textit{categories} (e.g, ``Java compilation'').

\subsubsection{Maven}

A failing build is caused by a thrown Java exception, which is present in the log. Therefore, we decided to use exception class names as a Maven error types, with the ``Exception'' suffix removed for brevity.

However, the exception can be chained. For example, a \texttt{LifecycleExecu\-tionException} can be caused by a \texttt{CompilationFailureException}, which can be caused by another one, etc. Fortunately, Maven logs contain a URL of a web page describing the relevant exception. We extracted the exception class name from this URL. In cases when multiple URLs were present, we extracted the class name from the last URL since it usually contained the most specific exception.

Some exception class types, e.g. \texttt{MojoExecutionExcep\-tion}, are too vague. In such cases, we searched for a presence of the pattern ``Failed to execute goal\dots'' and extracted the Maven plugin name which caused the failure, for instance, \texttt{maven-ja\-vadoc-plugin}. We then labeled the project with an error type in the form of ``ExceptionClass:plugin-name'' (see Table~\ref{t:maven} for examples).

\subsubsection{Gradle}

For Gradle, the thrown exception class was considered too. Since Gradle does not include a URL in its logs like Maven, we analyzed the printed Java exception ourselves. The situation was more complicated because Gradle sometimes utilizes multi-cause exceptions, when one exception is caused by more than one other exceptions, forming ``exception trees''. Assigning multiple error types to one build would unnecessarily complicate further analysis. Therefore, after a manual inspection of a subset of Gradle logs, we decided to use the following method to find the most relevant exception: If it was a simple (unchained) exception, it was selected. Otherwise, the first most direct cause of the root exception was selected. For example, in the exception tree ``A caused by B and C (caused by D)'', we selected the exception B.

Two exception classes were too vague, so we enriched the error type with the name of the failed task in these cases. Finally, for the \texttt{PluginApplicationExcep\-tion}, the failing plugin name was appended.

\subsubsection{Ant}

Since Ant throws the same exception (\texttt{BuildException}) for all possible build failure reasons, it is useless for error classification. However, Ant prints names of individual targets as it executes them. Therefore, we were able to extract the last executed target -- i.e., the failing one. The error type is in the form ``Target:target-name'' in this case. If the build failed before even one target was started, we analyzed the log for a presence of frequently occurring natural language patterns, determined by a manual inspection of the remaining logs.

\subsubsection{Categorization of Error Types}

In total, there were 292 different error types. Our goal was to categorize them, using human-readable names. Since this process was performed manually and the number of error types was large, we decided to categorize only more frequently occurring error types. While just 84 error types were categorized, they represented more than 86.5\% of build failures.

The categorization was performed by one of the authors. He discussed the documentation, inspected sample logs, and searched web forums to seek help with the categorization. When he was unsure, he left the error type uncategorized.

\subsection{Results}

First, we will present error types for individual build tools. Next, a overview of error categories will be presented.

\subsection{Error Types}

The most frequently occurring Maven error types are displayed in Table~\ref{t:maven}. A very large portion of Maven builds ends with the ``DependencyResolution'' error. This includes temporary and permanent network problems of the servers hosting the dependencies, authorization errors, missing files on servers due to reasons like discontinued dependency hosting or removed older versions, invalid POM files, and many others.

\begin{table}
\centering
\caption{The most frequent Maven error types}
\label{t:maven}
\begin{tabular}{lr}
\hline
\textbf{Error type} & \textbf{\%} \\ \hline
DependencyResolution & 39.11 \\
MojoFailure:maven-compiler-plugin & 18.16 \\
UnresolvableModel & 10.59 \\
MojoExecution:maven-javadoc-plugin & 8.17 \\
ProjectBuilding & 2.47 \\
PluginResolution & 2.41 \\
MojoExecution:git-commit-id-plugin & 1.70 \\
PluginManager & 1.65 \\
MojoExecution:maven-antrun-plugin & 1.37 \\
MojoExecution:exec-maven-plugin & 1.10 \\
MojoExecution:maven-enforcer-plugin & 1.04 \\
AetherClassNotFound & 0.77 \\ \hline
\end{tabular}
\end{table}

For the most frequent Gradle error types, see Table~\ref{t:gradle}. At the top, there is a ``CompilationFailed'' error which includes traditional Java compilation errors like undefined symbols, missing packages, incompatible types, etc.

\begin{table}
\centering
\caption{The most frequent Gradle error types}
\label{t:gradle}
\begin{tabular}{lr}
\hline
\textbf{Error type} & \textbf{\%} \\ \hline
CompilationFailed & 22.36 \\
MissingProperty & 11.59 \\
PluginApplication:ShadowJavaPlugin & 7.66 \\
Gradle:javadoc & 5.59 \\
InvalidUserData & 5.18 \\
ModuleVersionNotFound & 3.93 \\
Exec & 3.52 \\
ModuleVersionResolve & 3.52 \\
Resolve & 2.69 \\
IllegalArgument & 2.48 \\
CustomMessageMissingMethod & 2.07 \\
MultipleCompilationErrors & 1.86 \\ \hline
\end{tabular}
\end{table}

Table~\ref{t:ant} displays Ant error types. The most frequently failing target is ``compile''.

\begin{table}
\centering
\caption{The most frequent Ant error types}
\label{t:ant}
\begin{tabular}{lr}
\hline
\textbf{Error type} & \textbf{\%} \\ \hline
Target:compile & 21.83 \\
CannotImport & 17.03 \\
Target:-do-compile & 12.23 \\
Target:build & 3.28 \\
Target:build-project & 2.84 \\
unknown & 2.84 \\
Target:compile-import-shared & 2.62 \\
Target:-init-check & 1.75 \\
Target:jar & 1.75 \\
MissingPath & 1.53 \\
Target:-do-init & 1.53 \\
TaskdefNotFound & 1.31 \\ \hline
\end{tabular}
\end{table}

\begin{figure*}
\centering
\includegraphics[width=0.7\linewidth]{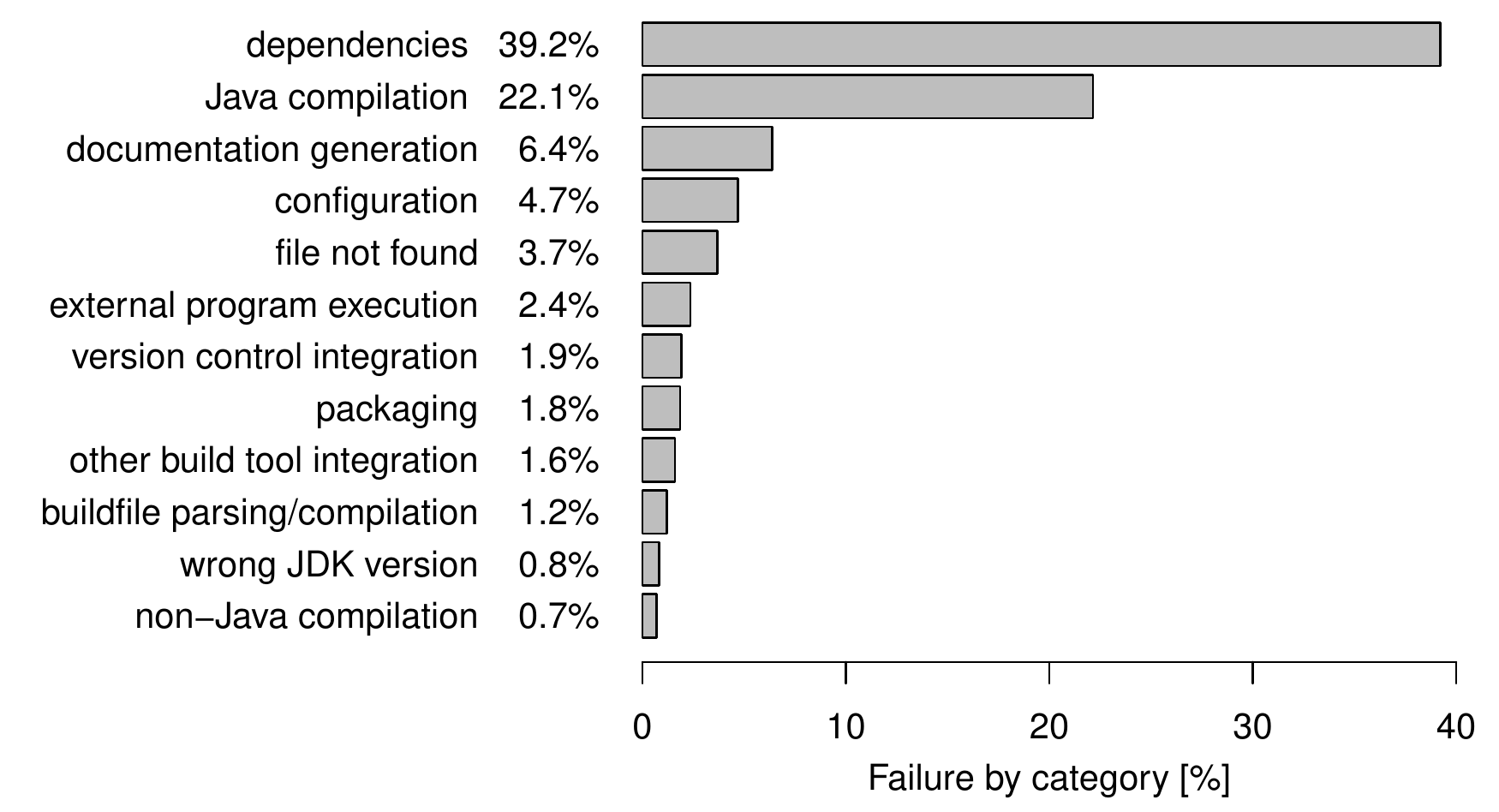}
\caption{The error categories}
\label{f:categories}
\end{figure*}

\subsection{Error Categories}

Error categories combined for all build tools are shown in Figure~\ref{f:categories}. Notably the largest portion of builds are dependency-related (39.2\%). This is caused by Maven builds to a large degree -- more than a half of failed Maven builds end with various dependency-related exceptions.

Compilation of Java source code caused 22.1\% of failures. Note that some compilation failures might be a consequence of missing dependencies \cite{Seo14programmers}.

Surprisingly, 6.4\% of failures occurred during documentation generation. Some builds failed because of missing or invalid configuration properties (4.7\%) and missing files or directories (3.7\%).

\section{Relation of Build Results to Project Properties}
\label{s:relation}

We hypothesize there is some form of association between the build result of a project and its properties. Specifically, we consider the project's build tool, size, popularity, age and last update recency.

\begin{itemize}
\item \textbf{H1:} The probability that a build fails depends on the build tool which the project uses.
\item \textbf{H2:} Builds of larger projects fail more likely than that of smaller ones. Project size was measured by a file count, as suggested by \citet{McIntosh15large}.
\item \textbf{H3:} Failing project are less popular, in terms of ``stars'' received on GitHub.
\item \textbf{H4:} Older projects, considering creation dates, fail more likely.
\item \textbf{H5:} More recently updated projects have a higher probability of a passing build.
\end{itemize}

\subsection{Method}

The build status is a nominal variable with two possible values: ``pass'' and ``fail''. The build tool is a nominal variable, while all other project properties are numeric (interval).

To be usable in statistical tests, dates must be converted to numeric values. We decided to use relative measures for readability reasons. Project age was measured as a number of days between the creation date of a particular project and of the newest project. A similar measure was used for recency -- using the ``last updated'' dates.

To determine whether the dependence of the build status and build tool is statistically significant, we used the Pearson's chi-squared test. To assess the dependence of build status and all other properties, a two-sided Mann-Whithey U test was performed. A confidence level of 99\% was used (p-value should be \textless{} 0.01).

\subsection{Results}

\begin{figure}
\centering
\includegraphics[width=0.9\linewidth]{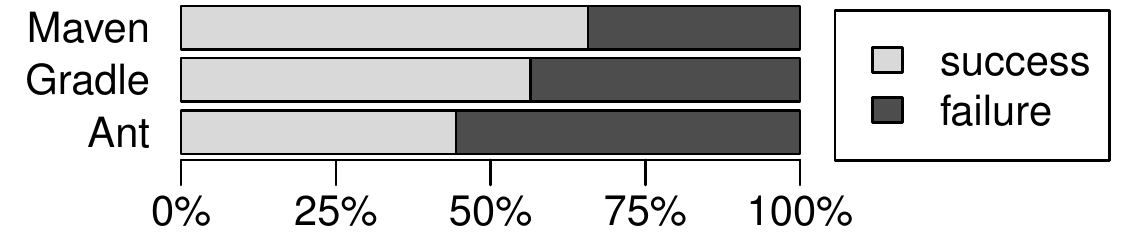}
\caption{The build status for individual build tools}
\label{f:tool-status}
\end{figure}

The chart of projects' build status for each build tool is in Figure~\ref{f:tool-status}. Maven builds tend to be the most successful (65.8\%), while Ant fails the most often (44.4\% success rate). The differences are statistically significant (p-value \textless{} 0.0001), confirming \textbf{H1}. Cram\'er's V is 0.146, which means small to medium effect size.

\begin{figure*}
\centering
\includegraphics[width=0.85\linewidth]{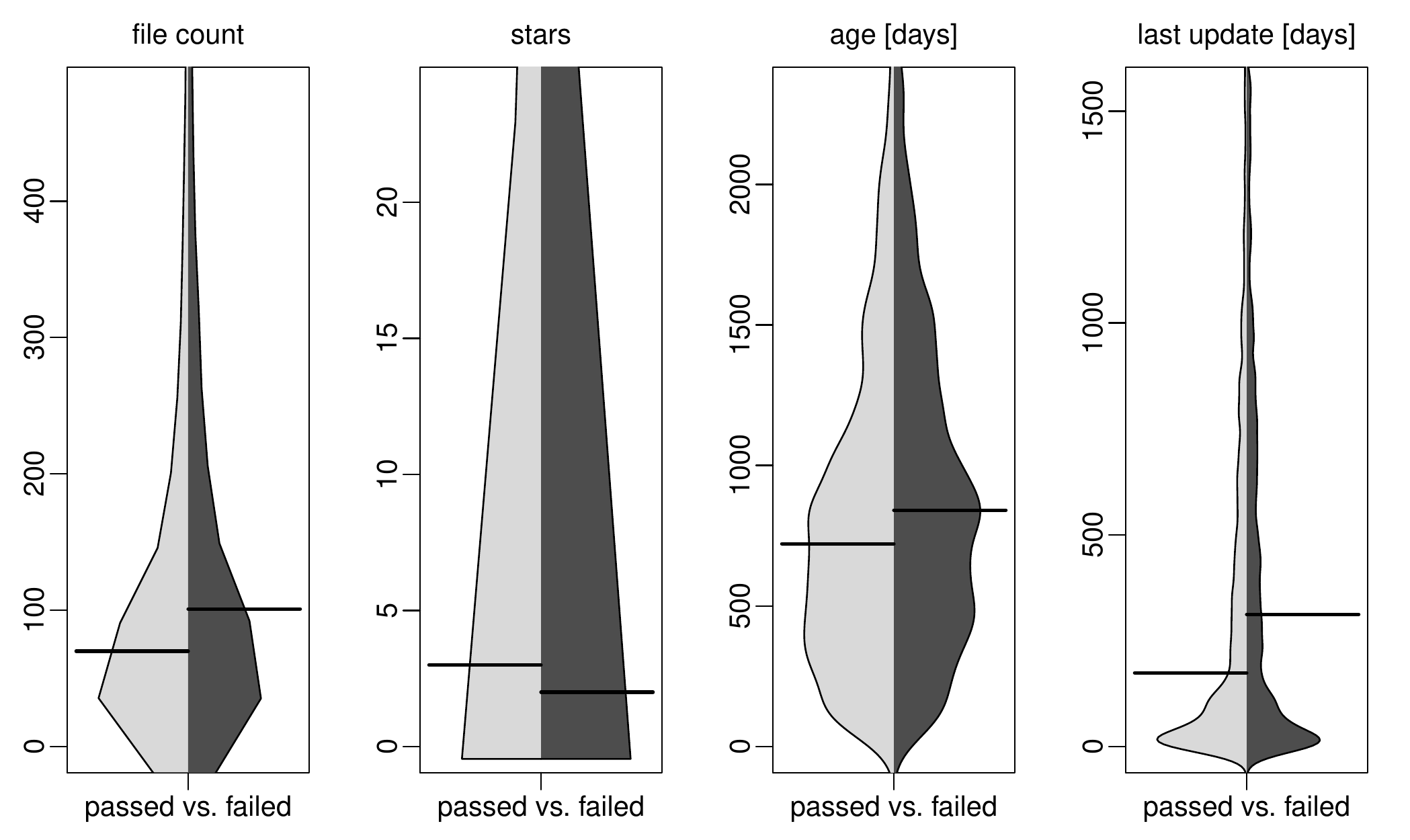}
\caption{Beanplots (distributions+medians) of project characteristics}
\label{f:assoc}
\end{figure*}

In Figure~\ref{f:assoc}, there are four beanplots depicting distributions and medians (horizontal lines) of file count, star count, age and update recency for projects with passed (left side) vs. failed (right side) builds. Outliers were trimmed from the display, but preserved in calculations.

Projects with a successful build had a median of 70 source files, while this number was 101 for the failing ones (a 44\% difference). This is natural -- larger projects are usually more complex and thus more error-prone. The result is statistically significant, with a p-value of less than $0.0001$, so \textbf{H2} is accepted.

There is a difference in the number of ``stars'' for the two groups -- a median of 3 vs. 2. The difference is not statistically significant (p = 0.0226).

When considering dates of creation (\textbf{H4}), there is \textasciitilde 17\% difference of age between passing (median \textasciitilde 722 days) and failing (841 days) builds. For a date of last update (\textbf{H5}), the situation is similar, but the difference is more striking: 174 vs. 312, i.e, a difference of 79\%. Both results are statistically significant (p \textless{} 0.0001). A possible explanation is that build scripts require maintenance -- even if the project is untouched -- to deal with changes in build tools themselves. After inspecting the results for individual build tools, Maven builds showed the largest difference.

Note that this is only an association study -- it does not mean that simply selecting Maven will make your project less prone to build failures. There may be complex relationships among multiple factors. For example, Ant projects tend to be older and slightly larger than other.

\section{Discussion}
\label{s:discussion}

Some portion of the failed projects could be buildable if we manually read the README file and followed the instructions provided. We argue that automated builds are not fully automated if they require a programmer to manually download dependencies, edit configuration files or perform other steps.

Nevertheless, we tried to manually download and build 3 random failing projects, using instructions provided in the README files (if present). If errors occurred anyway, we tried to correct them.

The first build failed even after we followed the instructions. It was successfully build only after we downgraded Gradle -- the build system itself -- from version 2.14 to 2.10. This shows that some projects require specific older versions of tools. An interesting fact is that the project was last updated less than a year before the study execution, so the ``project rot'' can occur quite fast.

The second project, with a name ending with ``-core'', failed because of a missing JAR file. This file could be produced by manually downloading and building a related ``-utils'' project. In README, there was only a brief mention that the project depends on the ``utils'' package. It is questionable why automated dependency management was not used, since both projects were using Maven, which has a built-in support for this.

The third project did not mention any building instructions in its README. Its build failed because a ``snapshot'' version of a third-party dependency was used. This version was probably available in the Maven repository in the past, but now it is not. After we corrected the version to a final one in the Maven POM file, a compilation error ``package does not exist'' occurred because of another missing dependency, unrelated to the previous one. Although we were trying to edit the POM file to download this dependency, builds were always failing with various download errors. We even tried downloading necessary JARs manually, but without success.

\section{Threats to Validity}

First, our results are valid only for the presented project selection criteria. Further studies should extend the criteria, particularly to include Android applications/libraries, which are becoming increasingly popular. Furthermore, about 80\% of GitHub repositories do not contain a license file\footnote{\url{http://github.com/blog/1964-open-source-license-usage-on-github-com}} -- including them could affect the results. Since analyzing GitHub alone has its disadvantages \cite{Kalliamvakou14promises}, we could also explore software forges other than GitHub.

The project's primary programming language was determined using the GitHub API, which uses a mapping of file extensions to languages, and returns a language in which the most bytes are written. This means the projects could utilize also other languages. However, we consider the presence of a Java build script in the project root directory a sign that the project (or its Java part) is buildable using Java tools. Furthermore, we excluded projects using JNI (although exclusion patterns for Android and JNI might not be faultless). Finally, only 2.4\% of failures were caused by external program execution, and even only some portion of them were because of an absence of a specific tool.

The selection of tools in our virtual environment might not represent a typical developer's setup. Nevertheless, they represent a minimal toolchain to download further tools and libraries as dependencies if necessary.

For Gradle and Maven, tests were excluded from the build process. For Ant, there is no command-line switch available for test exclusion, so they could potentially run. Our goal was that a single, universal command should be executed, which should just produce an output archive, without performing other activities like testing and deployment. The ad-hoc nature of Ant makes it difficult, so we resorted to the nearest behavior possible -- similar to what a real developer could do without modifying build scripts. Overall, less than 0.4\% of failures are test-related, which makes this threat negligible.

Some build tools cache downloaded dependencies. Our script does not clean the cache directories between individual builds. While this could in theory decrease internal validity, external validity is strengthened -- the programmer does not clean them often either.

Not all error types were categorized, so there may exist other categories, and the actual percentages of existing ones can slightly differ. However, 86.5\% of failed builds correspond to categorized error types. A similar approach was used by \citet{Seo14programmers} -- they covered about 90\% of their dataset.

Since Ant target names are generally free-form, they might not be the best categorization criterion. Using a custom logger, like in \cite{McIntosh10evolution}, we could extract task names to obtain more precise results.

Although categorization of error types was performed manually and solely by one researcher, this study is fully replicable and the analytical part is fully reproducible. The Docker image, scripts for analysis, and other materials are available at \url{http://sulir.github.io/build-study}.

\section{Related Work}

In this section, we will review some works related to build systems, with a focus on failure analysis.

\citet{Kerzazi14why} found that 18\% of automated builds of an industrial web application failed during a period of 6 months. In an exploratory study by \citet{Neitsch12build}, 4 of the 5 selected multilanguage Ubuntu packages could not be built or rebuilt without intervention. While these studies thoroughly investigated a small number of systems, our study takes a quantitative approach: it encompassed over 7,000 various projects.

A study \cite{Seo14programmers} conducted at Google showed that 30\% of developers' Java builds executed on a centralized build server failed. In contrast to our study, they used own proprietary build system, analyzed only compiler messages instead of full build system logs, and studied multiple builds of the same systems over time.

According to \citet{Beller16oops}, 59\% of broken builds on a continuous integration server Travis CI were caused by a test failure. Our study complements these results by examining failure reasons other than unsuccessful tests.

\citet{Vasilescu14continuous} described association of various project characteristics with the success of their build. They analyzed existing builds on a continuous integration server, while we created a virtual environment simulating a developer's local system.

\citet{McIntosh15large} performed a study of build systems on a large number of projects. Some of their findings are consistent with ours -- for example, that Maven requires the largest build maintenance effort. They were not interested in build successes and failures, though.

Among other characteristics, \citet{McIntosh10evolution} studied the build-time length of open source projects. However, success/failure rates and failure reasons were not presented.

In a study by \citet{Hochstein11cost}, 11\% of regression test failures were due to failed builds. They investigated only one project, though.

\section{Conclusion}

In a virtual environment containing programming tools, we tried to automatically build more than 7,000 Java projects from GitHub, corresponding to the selection criteria. Answering \textbf{RQ1}, more than 38\% of these projects failed to build. Regarding \textbf{RQ2}, the most frequent errors were dependency-related, followed by Java compilation and documentation generation. To answer \textbf{RQ3}, the likelihood of a build failure is associated with the build tool used; we also found that larger, older and less recently updated projects fail more.

Regarding the methodology, we presented an example of a simulation study in software engineering. In real life, we observed interesting behavior on a small number of projects. In a virtual environment, we tried to simulate the behavior which would be performed by programmers (i.e., building the projects) on a large sample and observed the outcomes.

While we did not study closed-source industrial systems themselves, they are also affected by this study since they often incorporate open source libraries in various ways.

\section{Future Work}

Especially for novices, error messages are often unreadable \cite{Marceau11mind}. By determining what build error messages are the most common, we can focus on making them more programmer-friendly. For example, we can break a frequently occurring generic message into multiple specific ones.

An important challenge is a more precise specification of the build environment and build process. A viable example is the Travis CI configuration file ``.travis.yml'', present in about 18\% of our studied projects. Its current role is to specify a build environment and steps in a machine-readable way for use on a CI (continuous integration) server. Extending its scope to local (desktop) builds seems like an interesting idea.

External dependency management causes both maintenance effort \cite{McIntosh15large} and build failures. Creating a system automatically generating and updating a build configuration by analyzing libraries in the source code would be useful.

Although we manifested the severity of build failures, we did not provide much guidance how to avoid them. It would be very useful to create a list of recommendations for programmers and build maintainers.

Researchers are welcome to perform further empirical studies using our images and scripts (\url{http://sulir.github.io/build-study}). Extending the scope to C, C++ and other languages, performing a longitudinal study -- analyzing the history of successful and broken builds over time, or measuring build/rebuild time are only a few interesting future research directions.

\acks

This work was supported by project KEGA No. 019TUKE-4/2014 Integration of the Basic Theories of Software Engineering into Courses for Informatics Master Study Programmes at Technical Universities -- Proposal and Implementation.


\bibliographystyle{abbrvnat}


\bibliography{plateau}

\end{document}